\documentstyle[aas2pp4]{article}

\journalid{337}{03 June 1997}
 
\lefthead{van Altena et al.}
\righthead{Distance to Hyades Cluster}

\begin{document}

\title{The Distance to the Hyades Cluster\\
       Based on HST Fine Guidance Sensor Parallaxes}

\author{W. F. van Altena, C. -L. Lu\altaffilmark{1}, J. T. Lee, T. M. Girard, 
X. Guo, C.~P.~Deliyannis\altaffilmark{2}, I.~Platais 
and V. Kozhurina-Platais}
\affil{Yale Astronomy Department, New Haven, CT 05620}

\author{B. McArthur, G. F. Benedict, R. L. Duncombe,
 P. D. Hemenway\altaffilmark{3}, 
 W. H. Jefferys, J.~R.~King\altaffilmark{2,4}, E.~Nelan\altaffilmark{4},
 P. S. Shelus, D. Story, and A. Whipple\altaffilmark{5}}
\affil{University of Texas at Austin, Austin, TX 78712}

\author{O. G. Franz and L. Wasserman}
\affil{Lowell Observatory, Flagstaff, AZ 86001}

\author{L. W. Fredrick}
\affil{Astronomy Department, University of Virginia,
 Charlottesville, VA 22903}

\author{R. B. Hanson, A. R. Klemola and B. F. Jones}
\affil{Lick Observatory, Santa Cruz, CA 95064}

\author{R. M\'endez}
\affil{European Southern Observatory, Karl-Schwarzschild-Stra$\beta$e 2,
D-85748, Garching bei M\"unchen, Germany}

\author{W. -S. Tsay}
\affil{Institute of Astronomy, National Central University, 
Chun-Li, Taiwan 32054, ROC}

\and

\author{A. Bradley}
\affil{Allied Signal Corporation, P. O. Box 91, Annapolis Jct., MD 20701}

\altaffiltext{1}{Current address: Purple Mountain Observatory, Chinese Academy
of Sciences, 2 Beijing Xi lu, Nanjing, Jiangsu 210080, PROC}
\altaffiltext{2}{Hubble Fellow}
\altaffiltext{3}{Current address: Department of Physics, University of Rhode
Island, Providence 02912}
\altaffiltext{4}{Current address: Space Telescope Science Institute,
 3700 San Martin Drive, Baltimore, MD 21218}
\altaffiltext{5}{Current address: Allied Signal Corporation, P. O. Box 91,
Annapolis Jct., MD 20701}

\clearpage
\begin{abstract}
Trigonometric parallax observations made with the Hubble 
Space Telescope's Fine Guidance Sensor \#3 (HST FGS) of seven Hyades 
Cluster members in six fields of view have been analyzed along with
their proper motions to determine the distance to the cluster. 
Knowledge of the Cluster's convergent point and mean proper motion are critical
to the derivation of the distance to the center of the cluster.  
Depending on the choice 
of the proper-motion system, the derived cluster center distance 
varies by 9\%.  Adopting a reference
distance of $46.1~pc$ or $m-M=3.32$, which is derived
from the ground-based parallaxes in the General Catalogue of Trigonometric
Stellar Parallaxes (1995 edition), the FK5/PPM proper-motion system 
yields a distance 4\%
larger, while the Hanson (1975) system yields a distance 2\% smaller.
The HST FGS
parallaxes reported here yield either a 14\% or 5\% larger distance depending
on the choice of the proper-motion system.
Orbital parallaxes (Torres et al. 1997a, 1997b, 1997c) yield an average 
distance 4\% larger than the reference distance.
The variation in the distance derived from the HST data illustrates the 
importance of the proper-motion system and the individual proper motions to 
the derivation of the distance to the Hyades Cluster center, therefore
a full utilization of the HST FGS parallaxes awaits the 
establishment of an accurate and consistent proper-motion system.
\end{abstract}
 

 
\section{Introduction}

	The Hyades Cluster is the nearest rich star cluster to 
the sun and it provides us with, among other things, a 
benchmark for the determination of the distances to other star 
clusters through the technique of main-sequence fitting.  
While the Pleiades Cluster is sometimes used as a standard for 
this process due to its more ``normal'' metallicity, its three 
times greater distance leads to a more uncertain zero-point in 
the derived distance scale.  Through the determination of the 
absolute magnitudes of nearby Classical Cepheids in clusters 
with respect to the Hyades and/or the Pleiades, the Population 
I extragalactic distance scale is derived.  

	Through the early 1960's the accepted distance to the Hyades Cluster
was determined by deriving the Convergent Point from the cluster members'
proper motions as was done, for example by \cite{van52}.  The distance was
questioned by \cite{hod66} who found
it to be in conflict with the results derived from a number of secondary 
distance estimators.  
Redeterminations of the Convergent Point by \cite{han75} using
absolute proper motions, \cite{gun88} and \cite{gri88} using radial velocities
and the proper motions from \cite{han75},
and most recently, Schwan (1990, 1991) using FK5 and PPM proper motions
have led to a distance that is approximately 17\% larger than the earlier
accepted value of 40~$pc$.  Using accurate masses for the double-lined
eclipsing binary vB22 and an adopted mass-luminosity relation for field stars,
\cite{mcc82}, followed by Peterson and Solensky (1987, 1988) who used the slope
of the Hyades mass-luminosity relation, obtained
a distance to the cluster center of $47~pc$.
Torres et al. (1997a, 1997b, 1997c) have derived orbital parallaxes 
from a combination of radial velocity and astrometric observations
which lead to a distance to the cluster center of about
48~$pc$ using the \cite{sch91} convergent point solution. The ground-based 
trigonometric parallaxes listed in the new edition of
the Yale Parallax Catalogue (van Altena et al. 1995), hereafter
referred to as the YPC,
were recently analyzed by \cite{van97a} who found a distance of 46~$pc$.  
The YPC investigation included 100 stars and used a weighted mean of the
parallaxes without use of the proper motions. This result should not be too
much in error due to the large number of stars and full spatial coverage of the
cluster. A more detailed analysis of the YPC data is in preparation.

	In 1968, \cite{van73} prepared a list of very probable 
Hyades Cluster members suitable for parallax determination 
and distributed it to numerous observatories involved in the 
determination of trigonometric parallaxes.  Those stars were 
selected to have high-accuracy proper motions indicative of 
membership in the Hyades, UBV photometry that placed the stars 
close to the main-sequence or white-dwarf ridge lines in the 
color-magnitude diagram and selected against double stars\footnote{
Griffin, et al. (1985) list vA627 = J285 as a single-lined spectroscopic
binary with a period of 850 days indicating that our screening 
against binary stars was not entirely successful. That should not be 
an important factor in this paper, since we are limited here by the 
lack of a consistent proper-motion system.}, 
and were in the magnitude range 9 through 14, i.e. accurately 
observable with the parallax telescopes and detectors then in 
use.  Many of the high-weight parallaxes analyzed in the YPC 
study were the result of intensive observational efforts on 
the stars in that list.  In addition, they formed the basis of 
our 1972 Phase B proposal to determine the distance to the 
Hyades Cluster using what was then called the Large Space 
Telescope and the 1977 Phase CD proposal for the Hubble Space 
Telescope.  Due to the reallocation of overhead in the 
observing procedures and ground control experienced by all 
Guaranteed Time Observers, and especially by those using the 
Fine Guidance Sensors, the original list of 20 Hyades Cluster 
members was reduced to 7 main-sequence members in 6 fields, 
each observed 6 to 7 times ($N_{obs}$ in Table 1) instead of the 
originally planned 24 times.  As a consequence, what was to be 
a definitive determination of the Hyades Cluster distance is 
now only a ``teaser''.  Finally, by the time this paper is in 
print, we will have the first results from the Hipparcos 
Astrometric Satellite on their determination of the distance 
to the Hyades Cluster.

\section{Observations and Reductions}

	The observations of the seven stars in six fields (Table 
1) were made over a period of three years from October 1993 
through September 1996, each field being observed during one 
orbit with FGS 3 at times of maximum parallax factor (average 
absolute value = 0.97).  Also listed in Table 1 for each 
field, is the Name of the Hyades cluster member from van Altena (1966, 1969)
that is the principal target (vA627 is in 
the same field as vA622), additional cross-identifications, the number
of reference stars used, 
$N_{ref}$, and the unit weight error of the parallax and proper-motion
solution in $x$ and $y$ corrected for degrees of freedom.  
\begin{table} [h]
\scriptsize
\begin{center}
\begin{tabular}{rccrrrrr}
\multicolumn{8}{c}{Table 1. HST FGS observations.}\\
\hline \hline
vA$^1$ & Ha$^2$ & GH7$^3$ & BD or Os$^4$  & $N_{obs}$ &
 $N_{ref}$ & $\sigma_1(x)$ & $\sigma_1(y)$ \\
    &       &       &            &      &      & $mas$    & $mas$ \\ \hline
310~ &  312  &  196  &  +17$^\circ$ 715  &  7   &    6 &   1.6  & 4.0 \\
383~ &  378  &  212  &  Os 373         &  7   &    6 &   2.1  & 3.4 \\
472$^5$ &  420  &  228  &  +13$^\circ$ 685  &  6   &    4 &   1.2  & 3.3 \\
548~ &  472  &  241  &  +15$^\circ$ 634  &  7   &    5 &   1.2  & 2.6 \\
622~ &  505  &  249  &                 &  7   &    7 &   2.2  & 2.6 \\
627~ &  509  &  250  &  +17$^\circ$ 744  &  7   &    7 &   2.2  & 2.6 \\
645~ &  517  &  253  &  Os 749         &  6   &    5 &   1.7  & 3.6 \\
\hline
\multicolumn{8}{l}{
\makebox[24pt][l]{notes:}\makebox[15pt][c]{1)}vA
 = van Altena (1969);\makebox[15pt][c]{2)}Ha = Hanson (1975);} \\
\multicolumn{8}{l}{ \makebox[24pt][l]{ }\makebox[15pt][c]{3)}GH7 =
Giclas, et al. (1962);\makebox[15pt][c]{4)}Os = Osvalds (1954);} \\
\multicolumn{8}{l}{ \makebox[24pt][l]{ }\makebox[15pt][c]
{5)}vA472 = vB100 in van Bueren (1952).} \\
\end{tabular}
\end{center}
\end{table}
The observing procedures and corrections for coordinate drift and 
Optical Field Angle Distortion (OFAD) were similar to those 
outlined in \cite{ben94}.  Coordinate drift in FGS 
3 during an orbit can amount to several thousandths of an 
arcsecond ($mas$) and for that reason, the target star and at 
times a second star were observed at the beginning of the 
orbit, half way through measurement of the six (on average) 
reference stars and again at the end.  Changes in the position 
of the target star and/or the second star were interpreted as 
a drift in the coordinate system and interpolated corrections 
were made to the positions of all measured stars.  The drift 
was modeled as being linear in time, although a quadratic 
drift yielded similar results.  The OFAD for FGS 3 was 
developed by \cite{jef92} and the OFAD appropriate 
to each observation date was computed by McArthur.  Local 
deviations of the actual focal plane from that predicted by 
the OFAD exist at the $mas$ level, but these introduce noise 
into the solutions and not systematic errors.  A minor 
systematic deviation of the OFAD from the focal plane was 
detected in the $y$-coordinate, but since the observations were 
made at maximum parallax factor, the $y$-solutions are used only 
for the proper-motion determination and not for the parallax.  
Since the target stars were about four magnitudes brighter 
than the reference stars we have searched for a possible 
systematic error as a function of star brightness, the 
magnitude equation.  No magnitude equation has been found in 
either the OFAD or Long Term Stability tests which both have 
magnitude ranges similar to the Hyades observations, so we do 
not believe that a magnitude equation exists in the Hyades 
data.  Since we have on average only six reference stars in 
each field ($N_{ref}$ in Table 1) and they are all rather faint (14 
- 16th mag.), we are unable to conclusively test for the 
existence of a magnitude equation in the Hyades data.

	The solutions for relative parallax and proper motion 
were made with the Yale parallax program developed by \cite{aue78} 
modified for use with HST FGS 
observations.  Parallel solutions were made by McArthur with 
the completely different University of Texas Gaussfit program 
by \cite{mca94} and negligible differences in the 
derived relative parallaxes attributable to weighting and 
modeling schemes were obtained.  The results presented here 
are from the Yale program.

	Since the parallaxes and proper motions determined with 
the HST FGS are relative to the means of those quantities for 
the reference stars, it was necessary to determine the 
respective corrections to absolute parallax and proper motion.  
The corrections to absolute parallax were computed from a 
galactic model used to compute those corrections for the YPC 
as well as from spectrophotometric parallaxes for the individual 
reference stars.  The spectrophotometric parallaxes used 
spectra obtained by Deliyannis and King with the WIYN telescope's 
MOS/Hydra spectrograph and CCD photometry obtained by I. 
Platais with the CTIO 0.9-meter telescope.  The data were 
reduced by Lu, Lee and Kozhurina-Platais and are being 
prepared for publication.  The two approaches yielded average 
corrections to absolute parallax of +1.3 to +1.4~$mas$; we have 
used the individual corrections derived from the 
spectrophotometric parallaxes.  The corrections to absolute 
proper motion were derived from a new galactic structure and 
kinematic model developed by \cite{men97} and 
measurements made for this purpose of the reference stars by Hanson, Klemola 
and Jones of the Lick Observatory Northern Proper Motion 
plates.  The Lick NPM corrections in $mas/yr$
for the individual reference stars in right ascension and 
declination were respectively ($+6.9\pm2,~-1.2\pm2$), while for 
400 faint anonymous stars of the same magnitude range they 
obtained ($+4.2\pm2,~-3.0\pm2$).  M\'endez calculated from his 
galactic structure and kinematic model ($+3.7\pm0.4,~-5.6\pm0.4$).  
We have adopted the Lick NPM corrections for the individual 
reference stars, although the final results are not 
significantly changed if we use the M\'endez corrections.  The 
error estimates for the Lick NPM proper motions are dominated 
by the zero-point error of the galaxy proper motions.

\begin{table*}
\scriptsize
\begin{center}
\begin{tabular}{rrrrrrrrrrrrrrr}
\multicolumn{15}{c}{\normalsize Table 2. Derived data for the Hyades Cluster
 Members$^{\dag}$.} \\ \hline \hline
 & & & & & & & & & & & \multicolumn{2}{c}{Schwan} & \multicolumn{2}{c}{Gunn} \\
Name\makebox[20pt][r]{$\pi^{\ddag}$}
 &\multicolumn{2}{c}{$\alpha~~(1950)~~\delta$}
 & V & B-V & $\pi$ & $\sigma$ & $\mu_{\alpha}$ & $\sigma$ & $\mu_{\delta}$
 & $\sigma$ & D$_c$ & $\sigma$ & D$_c$ & $\sigma$ \\
\makebox[20pt][r]{type} & \makebox[15pt][r]{h}\makebox[15pt][r]{m}\makebox[20pt
][r]{s} & \makebox[15pt][r]{$\deg$}\makebox[15pt][r]{$'$}\makebox[15pt][r]{$''$}
 & \multicolumn{2}{c}{(mag)} & \multicolumn{2}{c}{($mas$)} & \multicolumn{2}
 {c}{($mas/yr$)}  & \multicolumn{2}{c}{($mas/yr$)} & \multicolumn{2}{c}{(pc)}
 & \multicolumn{2}{c}{(pc)} \\ \hline
vA310\makebox[20pt][r]{H} & \makebox[15pt][r]{4}\makebox[15pt][r]{21}\makebox[
20pt][r]{22.9} & \makebox[15pt][r]{17}\makebox[15pt][r]{53}\makebox[15pt][r]{21}
 & 9.99 & 1.05 & 15.4 & 0.9 & 105.0 & 0.9 & -14.1 & 0.8 & 59.0 & 3.5 &
54.3 & 3.2 \\
vA383\makebox[20pt][r]{H} & \makebox[15pt][r]{4}\makebox[15pt][r]{23}\makebox[
20pt][r]{14.1}& \makebox[15pt][r]{14}\makebox[15pt][r]{ 55}\makebox[15pt][r]{46}
 & 12.14 & 1.45 & 16.0 & 0.9 & 91.7 & 0.8 & -15.8 & 1.0 & 52.1 & 3.2 &
47.9 & 3.0 \\
vA472\makebox[20pt][r]{H} & \makebox[15pt][r]{4}\makebox[15pt][r]{25}\makebox[
20pt][r]{15.1} & \makebox[15pt][r]{13}\makebox[15pt][r]{ 45}\makebox[15pt][r]{
29} & 9.03 & 0.84 & 16.6 & 1.6 & 78.9 & 1.3 & -16.7 & 1.5 & 44.5 & 4.6 &
40.8 & 4.3 \\
vA548\makebox[20pt][r]{H} & \makebox[15pt][r]{4}\makebox[15pt][r]{26}\makebox[
20pt][r]{38.9} & \makebox[15pt][r]{16}\makebox[15pt][r]{ 8}\makebox[15pt][r]{12}
 & 10.32 & 1.17 & 16.8 & 0.3 & 98.7 & 0.4 & -15.5 & 0.3 & 53.9 & 1.5 &
49.5 & 1.4 \\
vA622\makebox[20pt][r]{H} & \makebox[15pt][r]{4}\makebox[15pt][r]{28}\makebox[
20pt][r]{35.2} & \makebox[15pt][r]{17}\makebox[15pt][r]{ 36}\makebox[15pt][r]{
46} & 11.85 & 1.44&  21.6 & 1.1 & 99.6 & 1.0 & -26.1 & 1.3 & 43.4 & 2.4 &
39.9 & 2.2 \\
vA627\makebox[20pt][r]{H} & \makebox[15pt][r]{4}\makebox[15pt][r]{28}\makebox[
20pt][r]{43.2} & \makebox[15pt][r]{17}\makebox[15pt][r]{ 36}\makebox[15pt][r]{
15} & 9.55 & 0.98 & 16.5 & 0.9 & 106.6 & 1.1 & -16.2 & 3.0 & 58.9 & 3.5 & 
54.2 & 3.2 \\
vA645\makebox[20pt][r]{H} & \makebox[15pt][r]{4}\makebox[15pt][r]{29}\makebox[
20pt][r]{1.2} & \makebox[15pt][r]{15}\makebox[15pt][r]{ 23}\makebox[15pt][r]{38}
 & 11.05 & 1.28 & 15.7 & 1.2 & 102.5 & 1.4 & -14.0 & 1.4 & 60.8 & 4.8
& 55.9 & 4.4 \\ \hline
Wtd mean & & & & & & & & & & & 52.5 & 1.0 & 48.3 & 0.9 \\ \\

vB24\makebox[20pt][r]{O} & \makebox[15pt][r]{4}\makebox[15pt][r]{15}\makebox[
20pt][r]{25.4} & \makebox[15pt][r]{21}\makebox[15pt][r]{ 27}\makebox[15pt][r]{
31} & 5.87 & 0.24 & 17.9 & 0.6 & 101.4 & 0.9 & -36.6 & 1.0 & 49.1 & 1.6 & & \\
vB57\makebox[20pt][r]{O} & \makebox[15pt][r]{4}\makebox[15pt][r]{22}\makebox[
20pt][r]{45.8} & \makebox[15pt][r]{15}\makebox[15pt][r]{ 45}\makebox[15pt][r]{
42} & 7.05 & 0.49 & 21.4 & 0.7 & 105.0 & 3.1 & -25.3 & 3.5 & 44.8 & 1.9 & & \\
vB72\makebox[20pt][r]{O} & \makebox[15pt][r]{4}\makebox[15pt][r]{25}\makebox[
20pt][r]{48.2} & \makebox[15pt][r]{15}\makebox[15pt][r]{ 49}\makebox[15pt][r]{
42} & 3.74 & 0.18 & 21.2 & 0.8 & 111.7 & 1.6 & -25.5 & 1.8 & 49.0 & 1.9 & 
& \\ \hline
Wtd mean & & & & & & & & & & & 47.8 & 0.9 & & \\ \\

vB24\makebox[20pt][r]{G} & \makebox[15pt][r]{4}\makebox[15pt][r]{15}\makebox[
20pt][r]{25.4} & \makebox[15pt][r]{21}\makebox[15pt][r]{ 27}\makebox[15pt][r]{
31} & 5.87 & 0.24 & 19.4 & 1.1 & 96.8 & 0.7 & -36.0 & 0.4 & 43.5 & 2.5 & & \\
\hline
\multicolumn{13}{l}{\small \makebox[12pt][c]{\dag}For
   cross-identifications see Table 1.} \\
\multicolumn{13}{l}{\small \makebox[12pt][c]{\ddag}$\pi$ type definitions:} \\
\multicolumn{13}{l}{\small \makebox[24pt][r]{H} = HST trigonometric
parallax from this study.} \\
\multicolumn{13}{l}{\small \makebox[24pt][r]{O} = orbital
parallax from Torres et al. (1997a, 1997b, 1997c).} \\
\multicolumn{13}{l}{\small \makebox[24pt][r]{G} = ground-based
parallax from \cite{gat92}.} \\
\end{tabular}
\end{center}
\end{table*}

\section{Distance to the Cluster Center}

	The convergent point for the cluster was calculated by 
\cite{sch91} from 145 high-accuracy FK5 and PPM 
proper motions.  Using a subset of 62 stars found to lie within 4~$pc$ of 
the cluster center, he found a convergent point for the cluster 
($\alpha=97\fdg68\pm0\fdg42,~\delta=5\fdg98\pm0\fdg18$), 
a cluster center ($\alpha=65\fdg59,~\delta=16\fdg27$) and a 
distance of 47.9~$pc$.  Torres, et al (1997c) calculated the mean 
proper motion at the cluster center from 53 of the 62 stars as 
$\mu_c=113.1\pm0.7~mas/yr$. \cite{gun88} derived a slightly different 
convergent point
($\alpha=98\fdg2\pm1\fdg1,~\delta=6\fdg1\pm1\fdg0$) based
on the radial velocities determined by \cite{gri88} and the bulk proper motion
of the Hyades derived from the absolute proper motions of 59 stars from
\cite{han75}. Combined with their cluster center 
($\alpha=66\fdg15,~\delta=16\fdg65$), they obtained a distance to the
cluster center of $45.4\pm1.2~pc$.
We can calculate the distance of the Hyades cluster 
center, $D_c$, for each star observed with the HST FGS from,
\begin{equation}
	D_c = \pi^{-1}(\frac{sin\lambda_c}{sin\lambda})(\frac{\mu}{\mu_c})
\end{equation}
where the subscript $c$ refers to the cluster center, $\pi$ is the 
absolute parallax derived here for each star, $\lambda$ is the angular 
distance of the star from the convergent point on a great circle 
and $\mu$ is the absolute proper motion determined here along that great
circle.  The errors of the individual estimates of the cluster center 
distance were derived from a propagation of the errors of the proper 
motions and parallaxes, as the accidental errors in $\lambda, \lambda_c$, 
and $\mu_c$ do not contribute significantly to the total error.
Systematic errors in $\lambda_c$ and especially in $\mu_c$ do however 
have a {\it very} important effect.

In Table 2 we list the equatorial coordinates for the 
equinox 1950 from Hanson~(1975) for the first seven stars 
and from the PPM Catalogue for the remainder. Also given are the 
magnitudes and colors, absolute parallaxes, 
proper motions and their standard errors and the derived 
distance to the cluster center and its standard error. The latter two 
quantities are listed for both the \cite{sch91} and \cite{gun88} cluster 
parameters. The first part of the table lists the seven stars measured in the 
HST FGS  parallax program, while the second part lists the 
stars (vB24, vB57 and vB72) for which orbital parallaxes have been 
derived by Torres et al. (1997a, 1997b, 1997c), and the third part lists a 
trigonometric parallax 
and proper motion derived by \cite{gat92} for vB24 that was inadvertently 
omitted from the YPC.
The weighted mean distances of the cluster center are listed after each of the
first two sections along with their formal errors.

The various Hyades cluster distances are summarized in Table 3 along 
with their respective errors.  Internal errors are defined as the formal 
propagation of the parallax and proper motion errors into the error of the 
mean, while external errors are based on the dispersion of the individually 
derived cluster center distances.  
\begin{table} [h]
\scriptsize
\begin{center}
\begin{tabular}{lrrrr}
\multicolumn{5}{c}{Table 3. Hyades Cluster center distances (units: $pc$).} \\
\hline \hline
 & \multicolumn{2}{c}{\tiny Average and Internal Error}
 & \multicolumn{2}{c}{\tiny Weighted Mean and External Error} \\
 & \multicolumn{1}{c}{HST} & \multicolumn{1}{c}{Orbital}
 & \multicolumn{1}{c}{HST} & \multicolumn{1}{c}{Orbital} \\ \hline
Schwan & $53.2\pm1.0$ & $47.6\pm0.9$ & \makebox[12pt][r]{ }$52.5\pm2.7$
 & $47.8\pm1.4$\makebox[12pt][r]{ } \\
Gunn   & $48.9\pm0.9$ & & \makebox[12pt][r]{ }$48.3\pm2.0$ &  \\
\hline
\end{tabular}
\end{center}
\end{table}
As can be seen from a comparison of the 
HST distances based on the \cite{sch91} and \cite{gun88} solutions, the 
results depend critically on which convergent point solution is adopted 
with the dominant factor being the proper-motion system and the individual 
proper motions.  According to Eqn. (1), after the well-defined scaling 
due to differing angular distances from the convergent point, the distance 
of a star relative to the cluster center is given by the ratio of 
the star's proper motion to that of the cluster center.  
The cluster center distance is then derived from the scaled proper motion 
distance and the parallax of the star. The HST parallaxes yield 
a cluster center distance ($48.3\pm2.0~pc;~3.42\pm0.09~mag$) 
in agreement with the orbital parallaxes 
and the YPC for the the \cite{gun88} solution since the HST proper motions 
are small relative to the bulk proper motion of the Hyades center as derived 
by \cite{gun88} from the \cite{han75} proper motions and the HST parallaxes 
are small. In contrast, 
the smaller \cite{sch91} cluster center proper motion places the HST 
stars closer to the center and therefore the small HST parallaxes move the
center farther ($52.5\pm2.7~pc;~3.60\pm0.11~mag$) from the sun.  
This emphasizes the importance of the proper-motion system in determining the
distance to the Hyades cluster center when small data samples are used.
The distance derived from the orbital parallaxes 
is essentially independent of the convergent point solution ($1\%$ difference) 
since for consistency both must 
use the FK5/PPM proper motions and the cluster center proper motion derived 
from \cite{sch91}, as the three stars were either too bright for accurate 
measurement in Hanson's (1975) study, or were outside his field of view.

	It would be unwise to advocate any increase in the 
distance scale based on the HST FGS parallax results, but it 
should be noted that \cite{fea97} recommend an 
increase of $10\%\pm14\%$ based on Hipparcos parallax observations 
of the Classical Cepheids and \cite{rei97} suggests an increase 
of 5\% to 15\% based on Hipparcos parallaxes of Subdwarfs.  Once 
the Hipparcos parallaxes of Hyades cluster members are 
released and carefully analyzed we should have a clearer picture of the 
state of the distance scale and can then discuss the 
astronomical consequences of any revision\footnote{At 
the Hipparcos Venice 1997 meeting, Brown, et al. (1997) and Perryman, et al.
(1997) reported a distance to the Hyades Cluster center of 
$46.34\pm0.27~pc$, or $m-M = 3.33\pm0.01$ based on 134 stars.
The weighted average distance derived from the four Hipparcos stars in common
with our seven HST stars is $46.46\pm2.24~pc$, or $3.34\pm0.10~mag$. It appears
that there are systematic differences between these HST and the Hipparcos
parallaxes and proper motions, and we are continuing to investigate those
differences. The Hipparcos
distance is in excellent agreement with the YPC ground-based parallaxes 
reported by van Altena, et al (1997a) and indicates that no change in the 
distance scale is required.}.

	The Fourth Edition of the General Catalogue of 
Trigonometric Stellar Parallaxes in two volumes is available 
in printed form from the Yale Astronomy Department and in an abbreviated
electronic form from the Astronomical Data Centers.

\section{Acknowledgments}

The many authors of this paper would like to acknowledge the assistance 
of the STScI and GSFC staffs who assisted us in the preparation of numerous
versions of the Observing Proposal forms, and provided solutions for 
obstacles that were encountered during the 
observations; we would not have been able to complete the 
observations without their invaluable help.  In particular we 
would like to acknowledge the collaboration of L. Nagel, D. 
Taylor and P. Stanley.  In addition we would like to 
acknowledge the engineers and scientists at the Hughes Danbury 
Optical Systems who designed the FGS and supported us in the 
calibration and observations.  In particular we would like to 
mention L. Abramowicz-Reed, C. Ftaclas and T. Facey.  Finally, we would 
like to thank the referee of this paper, D. Latham, for his many helpful 
suggestions which significantly improved the manuscript.

	This research was supported in part by grants from NASA 
to the HST GTO Astrometry Science Team.  The preparation of 
the YPC was supported in part by grants from the NSF. Deliyannis and King also 
acknowledge support by NASA through grant number HF-1042.01-93A
and HF-1046.01-93A from the Space 
Telescope Science Institute, which is operated by the Association of 
Universities for Research in Astronomy, Inc., under NASA contract 
No. NAS5-26555.


\end{document}